\documentclass[12pt]{article}

\def \la {{\langle}}
\def \ra {{\rangle}}

\newcommand{\be}{\begin{equation}}
\newcommand{\ee}{\end{equation}}
\newcommand{\beqar}{\begin{eqnarray}}
\newcommand{\eeqar}{\end{eqnarray}}

\newcommand{\Tr}{{\rm Tr}}
\newcommand{\half}{{\frac{1}{2}}}

\expandafter\ifx\csname mathbbm\endcsname\relax

\else

\fi
\begin{document}
\begin{titlepage}
\begin{flushleft}  
       \hfill                      CCNY-HEP-01-02\\
       \hfill                      RU-01-4-B\\
       \hfill                      {\tt hep-th/0103013}\\
       \hfill                       March 2001\\
\end{flushleft}
\vspace*{3mm}
\begin{center}
{\LARGE Quantum Hall states as matrix Chern-Simons theory
\\}
\vspace*{12mm}
{\large Alexios P. Polychronakos\footnote{On leave from Theoretical 
Physics Dept., Uppsala University, 751 08 Sweden and Dept.~of Physics, 
University of Ioannina, 45110 Greece; E-mail: poly@teorfys.uu.se} \\
\vspace*{5mm}
{\em Physics Department, City College of the CUNY \\
New York, NY 10031, USA}\\
\vspace*{5mm}
and \\
\vspace*{5mm}
{\em Physics Department, The Rockefeller University \\
New York, NY 10021, USA\/}\\}
\vspace*{15mm}
\end{center}

\begin{abstract}
We propose a finite Chern-Simons matrix model on the plane
as an effective description of fractional quantum Hall fluids
of finite extent. The quantization of the inverse filling
fraction and of the quasiparticle number is shown to arise
quantum mechanically and to agree with Laughlin theory.
We also point out the effective equivalence of this model,
and therefore of the quantum Hall system, with the Calogero
model.
\end{abstract}

\end{titlepage}

\section{Introduction}

Chern-Simons actions have appeared in various contexts as
topological terms in the action for gauge fields in odd
(spacetime) dimensions \cite{DJT}. Their generalization to
noncommutative spaces have also been considered both in
the star-product formulation \cite{CF,Kr} and the operator 
formulation \cite{PolCS}, and have received a lot of attention
in the context of string and $D$-brane physics, induced
actions, vortex and nonperturbative solutions etc. 
\cite{Chu}-\cite{BKSY}.

The planar coordinates of quantum particles in the lowest 
Landau level of a constant magnetic field provide a well-known 
and natural realization of noncommutative space \cite{DuJT}.
This connection had so far not proven to be of much relevance
to noncommutative field theory. 
Recently, however, Susskind proposed that noncommutative
Chern-Simons theory on the plane may provide a description
of the (fractional) quantum Hall fluid \cite{Suss}, and 
specifically of the Laughlin states \cite{Laug}. In this, 
noncommutative Chern-Simons theory finds a new context 
and acquires new interest. Connections between D-brane
physics and the quantum Hall effect were also previously
examined in \cite{BST}-\cite{CCMM}.

Chern-Simons theory on the plane necessarily describes a 
spatially infinite
quantum Hall system. The space noncommutativity condition requires
an infinite dimensional Hilbert space, and the fields of the theory
become operators in this space. It is, however, of interest to
also describe finite systems, of limited spatial extent with a
finite number of electrons. Such systems form quantum Hall
`droplet' states and exhibit boundary excitations.

The purpose of this paper is to provide such a description. 
An appropriate finite matrix model will be proposed and shown 
to reproduce the basic features of the quantum Hall states. 
The quantization of the model will be performed and shown to imply
a quantization of the filling fraction, similar to the 
quantization of the planar noncommutative Chern-Simons 
theory \cite{NP} (see also \cite{BLP}), which in turn 
implies that the underlying 
particles (electrons) obey integral statistics.
Further, the charge of elementary excitations inside the fluid
(quasiholes) will be calculated and shown to be quantized in
units of the filling fraction, in accordance with Laughlin theory.
It will also be pointed out that this model is essentially equivalent 
to the Calogero model \cite{Cal}-\cite{PolLH} a well-known 
integrable system whose connection to fractional statistics
\cite{PolFS,LM} and anyons \cite{PolAN}-\cite{Ouv}
has been established in different contexts. Finally, some remarks
on the expected phase transition of the
quantum Hall state to a Wigner crystal will be offered.

\section{Chern-Simons theory on the noncommutative plane
and quantum Hall states}

\subsection{Review of Chern-Simons theory}

We briefly review the basic features of Chern-Simons (CS) theory 
on a noncommutative plane and a commutative time and its connection
with quantum Hall states \cite{Suss}. 

The system to be described consists of $N \to \infty$ electrons 
on the plane in an external constant magnetic field $B$ 
(we take their charge $e=1$).
Their coordinates are globally parametrized in a fuzzy way by two
(infinite) hermitian `matrices' $X_a$, $a=1,2$, that is, by two
operators on an infinite Hilbert space. Their average density is 
$\rho_0 = 1/2\pi \theta$. The action is the analog of the gauge 
action of particles in a magnetic field:
\be
S = \int dt\, \frac{B}{2} \, \Tr \left\{ \epsilon_{ab} ({\dot X}_a 
+i [A_0 , X_a ]) X_b + 2\theta A_0 \right\}
\label{CSX}
\ee
with $\Tr$ representing (matrix) trace over the Hilbert space and
$[.,.]$ representing matrix commutators. The above has the form of a
noncommutative CS action, which in operator form reads \cite{PolCS}
\be
S = \lambda \int dt\, 2\pi \theta\, \Tr \left\{ \frac{2}{3} 
\epsilon_{\mu \nu \rho} D_\mu D_\nu D_\rho + \frac{2}{\theta} 
A_0 \right\}
\label{CSD}
\ee
The `coordinates' $X_a$, are the analogs of the covariant derivative 
operators, $X_a \sim \theta D_a$, and $D_0 = -i\partial_t + A_0$. 
The time component of the gauge field ensures gauge invariance, 
its equation of motion imposing the Gauss law constraint
\be
[ X_1 , X_2 ] = i \theta
\label{XX}
\ee
The term in the action linear in $A_0$ is crucial to ensure that 
the noncommutative vacuum (\ref{XX}) is a solution of the equations, 
with the noncommutativity parameter $1/\theta$ being 
essentially the particle density. The coefficient of the CS 
term $\lambda$ relates to $B$ as
\be
\lambda = \frac{B \theta}{4\pi} = \frac{1}{4\pi \nu}
\label{lambda}
\ee
with $\nu = 2\pi \rho_0 / B$ being the filling fraction.

Gauge transformations are conjugations of $X_a$ or $D_a$ by arbitrary
time-dependent unitary operators. In the quantum Hall context they
take the meaning of reshuffling the electrons. Equivalently, the
$X_a$ can be considered as coordinates of a two-dimensional fuzzy
membrane, $2\pi\theta$ playing the role of an area quantum and gauge 
transformations realizing area preserving diffeomorphisms.
The canonical conjugate of $X_1$ is $P_2 = B X_2$, and the 
generator of gauge transformations is
\be
G = -i B [ X_1, X_2 ] = B\theta = \frac{1}{\nu}
\label{Gauss}
\ee
by virtue of (\ref{XX}). Since gauge transformations are interpreted
as reshufflings of particles, the above has the interpretation of 
endowing the particles with quantum statistics of order $1/\nu$.

\subsection{Quasiparticle and quasihole states}

The classical equation (\ref{XX}) has a unique solution, modulo
gauge (unitary) transformations, namely the unique irreducible
representation of the Heisenberg algebra. So the classical theory
has a unique state, the vacuum.
Deviations from the vacuum (\ref{XX}) can be achieved by introducing
sources in the action \cite{Suss}. A localized source at the origin 
has a density
of the form $\rho = \rho_0 - q \delta^2 (x)$ in the continuous
(commutative) case, representing a point source of particle number
$-q$, that is, a hole of charge $q$ for $q>0$. The noncommutative 
analog of such a density is
\be
[ X_1 , X_2 ] = i \theta ( 1 + q | 0 \ra \la 0 | )
\label{q}
\ee
where $|n\ra$, $n=0,1,\dots$ is an oscillator basis for the
(matrix) Hilbert space, $|0\ra$ representing a state of minimal
spread at the origin. In the membrane picture the right-hand side
of (\ref{q}) corresponds to area and implies that the area quantum
at the origin has been increased to $2\pi\theta(1+q)$, therefore
piercing a hole of area $A = 2\pi\theta q$ and creating a particle 
deficit $q = \rho_0 A$. We shall call this a quasihole state.
(In \cite{Suss} these were called quasiparticle states; we shall 
use the opposite terminology here to conform with the standard
definition of particle number.) For $q>0$ we find the quasihole
solution of (\ref{q}) as 
\cite{Suss}
\be
X_1 + i X_2 = \sqrt{2\theta} \sum_{n=1}^\infty \sqrt{n+q}
\, | n-1 \ra \la n |
\label{Xqpos}
\ee
as was also found in \cite{Suss}.

The case of quasiparticles, $q<0$ is more interesting. 
Clearly the area quantum cannot be diminished below zero, and 
equations (\ref{q}) and (\ref{Xqpos}) cannot hold for $-q>1$. 
The correct equation is, instead,
\be
[ X_1 , X_2 ] = i \theta \left( 1 - \sum_{n=0}^{k-1} 
| n \ra \la n | - \epsilon | k \ra \la k | \right)
\label{qneg}
\ee
where $k$ and $\epsilon$ are the integer and fractional part
of the quasiparticle charge $-q$. The solution of (\ref{qneg}) is
\be
X_1 + i X_2 = \sum_{n=0}^{k-1} z_n | n \ra \la n | + 
\sqrt{2\theta} \sum_{n=k+1}^\infty \sqrt{n-k-\epsilon} 
\, | n-1 \ra \la n |
\label{Xneg}
\ee
(For $k=0$ the first sum in (\ref{qneg},\ref{Xneg}) drops.) 
In the membrane picture, $k$ quanta of the membrane have `peeled' 
and occupy positions $z_n = x_n + i y_n$ on the plane, while the
rest of the membrane has a deficit of area at the origin equal
to $2\pi \theta \epsilon$, leading to a charge surplus $\epsilon$.
Clearly the quanta are electrons that sit on top of the continuous
charge distribution. If we want all charge density to be concentrated
at the origin, we must choose all $z_n =0$.

We note that the above quasiparticle states for integer $q$ are
basically the same as the noncommutative solitons and flux tubes
found in noncommutative gauge theory \cite{GMS,PolFT,GN,HKL},
while the quasihole states have no analog.

Laughlin theory predicts that quasihole excitations in the 
quantum Hall state have their charge $-q$ quantized in integer units
of $\nu$, $q = m \nu$, with $m$ a positive integer. We see that 
the above discussion gives no hint of this quantization, while 
we see at least some indication of electron quantization in 
(\ref{qneg},\ref{Xneg}). Quasihole quantization will emerge in 
the quantum theory, as we shall see shortly.

\section{A model for finite number of electrons}
\subsection{The Chern-Simons finite matrix model}

We come now to the problem of describing quantum Hall states of
finite extent consisting of $N$ electrons. Obviously the coordinates 
$X_a$ now have to be represented by finite $N \times N$ matrices.
The action (\ref{CSX}), however, and the equation (\ref{XX}) to
which it leads, are inconsistent for finite matrices, and a modified
action must be written which still captures the physical features
of the quantum Hall system. Such an action in fact 
exists \cite{PolMM}. For our purposes we take
\be
S = \int dt \frac{B}{2} \Tr \left\{ \epsilon_{ab} ({\dot X}_a 
+i [A_0 , X_a ]) X_b + 2\theta A_0 - \omega X_a^2 \right\} 
+ \Psi^\dagger (i {\dot \Psi} - A_0 \Psi)
\label{CSPsi}
\ee
It has the same form as the planar CS action, but with two extra
terms. The first involves $\Psi$, a complex $N$-vector that 
transforms in the fundamental of the gauge group $U(N)$: 
\be
X_a \to U X_a U^{-1} ~,~~~\Psi \to U \Psi
\ee
Its action is a covariant kinetic term similar to a complex scalar
fermion. We shall, however, quantize it as a boson; this is
perfectly consistent, since there is no spatial kinetic term that
would lead to a negative Dirac sea and the usual inconsistencies
of first-order bosonic actions.

The term proportional to $\omega$ serves as a spatial regulator:
since we will be describing a finite number of electrons, there 
is nothing to keep them localized anywhere in the plane. 
We added a confining harmonic potential which serves as a `box' 
to keep the particles near the origin.

We can again impose the $A_0$ equation of motion as a Gauss
constraint and then put $A_0 =0$. In our case it reads
\be
G \equiv -iB [ X_1 , X_2 ] + \Psi \Psi^\dagger - B\theta =0
\label{G}
\ee
Taking the trace of the above equation gives
\be
\Psi^\dagger \Psi = NB \theta
\ee
The equation of motion for $\Psi$ in the $A_0 =0$ gauge is 
${\dot \Psi} =0$. So we can take it to be
\be
\Psi = \sqrt{NB} \, |v\ra
\ee
where $|v\ra$ is a constant vector of unit length. The traceless 
part of (\ref{G}) reads
\be
[ X_1 , X_2 ] = i\theta \left( 1 - N |v\ra \la v| \right)
\label{XXv}
\ee
This is similar to (\ref{XX}) for the infinite plane case,
with an extra projection operator. Using the residual gauge freedom
under time-independent unitary transformations, we can rotate
$|v\ra$ to the form $|v\ra = (0,\dots 0,1)$. The above commutator
then takes the form $i\theta \,{diag}\, (1,\dots, 1, 1-N)$ which
is the `minimal' deformation of the planar result (\ref{XX})
that has a vanishing trace. 

In the membrane picture, $\Psi$
is like a boundary term. Its role is to absorb the `anomaly' of the
commutator $[X_1 , X_2 ]$, very much like the case of a boundary
field theory required to absorb the anomaly of a bulk CS field theory.

The equations of motion for $X_a$ read
\be
{\dot X}_a + \omega \epsilon_{ab} X_b =0
\ee
This is just a matrix harmonic oscillator. It is solved by
\be
X_1 + i X_2 = e^{i\omega t} A
\label{Arot}\ee
where $A$ is any $N \times N$ matrix satisfying the constraint
\be
[ A, A^\dagger ] = 2\theta (1 - N |v\ra \la v| )
\label{AA}
\ee

The classical states of this theory are given by the set of
matrices $A = X_1 + i X_2$ satisfying (\ref{AA}) or (\ref{XXv}).
We can easily find them by choosing a basis in which one of
the $X$'s is diagonal, say, $X_1$. Then the commutator 
$[ X_1 , X_2 ]$ is purely off-diagonal and the components of
the vector $|v\ra$ must satisfy $|v_n |^2 = 1/N$. We can use
the residual $U(1)^N$ gauge freedom to choose the 
phases of $v_n$ so that $v_n = 1/\sqrt{N}$. So we get
\be
(X_1 )_{mn} = x_n \delta_{mn} ~,~~~
(X_2 )_{mn} = y_n \delta_{mn} + \frac{i\theta}
{x_m - x_n} (1-\delta_{mn} )
\label{Xsol}
\ee
The solution is parametrized by the $N$ eigenvalues if $X_1$, 
$x_n$, and the $N$ diagonal elements of $X_2$, $y_n$. 

\subsection{Quantum Hall `droplet' vacuum}

Not all solutions found above correspond to quantum Hall fluids. 
In fact, choosing all $x_n$ and $y_n$ much bigger than $\sqrt\theta$
and not too close to each other, both $X_1$ and $X_2$ become
almost diagonal; they represent $N$ electrons scattered in 
positions $(x_n , y_n )$ on the plane and performing rotational
motion around the origin with angular velocity $\omega$. This
is the familiar motion of charged particles in a magnetic field 
along lines of equal potential when their proper kinetic term
is negligible. Quantum Hall states will form when particles
coalesce near the origin, that is, for states of low energy. 

To find the ground state, we must minimize the potential
\be
V = \frac{B\omega}{2} \Tr (X_1^2 + X_2^2 ) = 
\frac{B\omega}{2} \Tr (A^\dagger A)
\ee
while imposing the constraint (\ref{XXv}) or (\ref{AA}). This
can be implemented with a matrix Lagrange multiplier $\Lambda$
and we obtain
\be
A = [\Lambda , A] ~,~~~~{\rm or}~~~ X_a = i \epsilon_{ab}
[ \Lambda , X_b ]
\label{AAL}
\ee
This is reminiscent of canonical commutation relations for a
quantum harmonic oscillator, with $\Lambda$ playing the role 
of the hamiltonian. We are led to the solution
\be
A = \sqrt{2\theta} \sum_{n=0}^{N-1} \sqrt{n} |n-1\ra \la n | 
~,~~~  \Lambda = \sum_{n=0}^{N-1} n |n\ra \la n | 
~,~~~ |v\ra = | N-1 \ra
\label{APN}
\ee
This is essentially a quantum harmonic oscillator and 
hamiltonian projected to the lowest $N$ energy eigenstates. 
It is easy to check that the above satisfies both 
(\ref{AA}) and (\ref{AAL}). 
Its physical interpretation is clear: 
it represents a circular quantum Hall `droplet' of radius 
~$\sqrt{2N\theta}$. Indeed, the radius-squared matrix 
coordinate $R^2$ is
\beqar
R^2 &=& X_1^2 + X_2^2 = A^\dagger A + \half [a,a^\dagger] \\
    &=& \sum_{n=0}^{N-2} \theta (2n+1) |n\ra \la n| +
\theta (N-1) |N-1 \ra \la N-1 |
\eeqar
The highest eigenvalue of $R^2$ is $(2N-1)\theta$. The particle 
density of this droplet is $\rho_0 = N/(\pi R^2) \sim 
1/(2\pi \theta)$ as in the infinite plane case. 

The matrices $X_a$ are known and can be explicitly diagonalized 
in this case. Their eigenvalues are given by the
zeros of the $N$-th Hermite polynomial (times $\sqrt{2\theta}$).
In the large-$N$ limit the distribution of these zeros obeys the 
famous Wigner semi-circle law, with radius $\sqrt N$. Since these
eigenvalues are interpreted as electron coordinates, this
confirms once more the fact that the electrons are evenly 
distributed on a circle of radius $~\sqrt{2N\theta}$.

\subsection{Excited states of the model}

Excitations of the classical ground state can now be considered.
Any perturbation of (\ref{APN}) in the form of (\ref{Xsol}) is,
of course, some excited state. We shall concentrate, however,
on two special types of excitations. 

The first is obtained by performing on $A,A^\dagger$ all
transformations generated by the infinitesimal transformation
\be
A' = A + \sum_{n=0}^{N-1} \epsilon_n (A^\dagger)^n
\label{dropex}
\ee
with $\epsilon_n$ infinitesimal complex parameters. The sum is
truncated to $N-1$ since $A^\dagger$ is an $N \times N$ matrix
and only its first $N$ powers are independent. It is
obvious that (\ref{AA}) remains invariant under the above
transformation and therefore also under the finite transformations
generated by repeated application of (\ref{dropex}). 

If $A,A^\dagger$
were true oscillator operators, these would be canonical (unitary)
transformations, that is, gauge transformations that would
leave the physical state invariant. For the finite $A,A^\dagger$
in (\ref{APN}), however, these are {\it not} unitary transformations
and generate a new state. To understand what is that new state,
examine what happens to the `border' of the circular quantum Hall
droplet under this transformation. This is defined by 
$A^\dagger A \sim 2N\theta$ (for large $N$). To find the new boundary
parametrize $A \sim \sqrt{2N\theta} e^{i\phi}$, with $\phi$ the 
polar angle on the plane and calculate $(A^\dagger A)'$. 
The new boundary in polar coordinates is
\be
R' (\phi ) = \sqrt{2N\theta} + \sum_{n=-N}^N c_n e^{in\phi} 
\label{newR}
\ee
where the coefficients $c_n$ are
\be
c_n = c_{-n}^* = \frac{R^n}{2} \epsilon_{n-1}
~~(n>0),~~~~c_0 =0
\label{cc}
\ee
This is an arbitrary area-preserving deformation of the
boundary of the droplet, truncated to the lowest $N$ 
Fourier modes. The above states are, therefore, 
arbitrary area-preserving boundary excitations of the
droplet \cite{Wen,IKS,CTZ}, appropriately truncated to reflect
the finite noncommutative nature of the system (the fact
that there are only $N$ electrons). 

Note that on the plane there is an infinity of 
area-preserving diffeomorphisms that produce a specific 
deformation of a given curve. From the droplet point of view,
however, these are all gauge equivalent since they deform
the outside of the droplet (which is empty) or the inside
of it (which is full and thus invariant). The finite theory
that we examine has actually broken this infinite gauge
freedom, since most of these canonical transformations of
$a,a^\dagger$ do not preserve the Gauss constraint (\ref{AA})
when applied on $A,A^\dagger$. The transformations (\ref{dropex})
pick a representative in this class which respects the constraint.

The second class of excitations are the analogs of quasihole and
quasiparticle states. States with a quasihole of charge $-q$ at 
the origin can be written quite explicitly in the form
\be
A = \sqrt{2\theta} \left( \sqrt{q} |N-1 \ra \la 0| + \sum_{n=1}^{N-1} 
\sqrt{n+q} |n-1\ra \la n| \right) ~,~~~ q>0
\label{qhole}
\ee
It can be verified that the eigenvalues of $A^\dagger A$ are
\be
(A^\dagger A)_n = 2\theta(n+q) ~,~~~ n=0,1,\dots N-1
\label{eigen}
\ee
so it represents a circular
droplet with a circular hole of area $2\pi\theta q$ at the origin,
that is, with a charge deficit $q$. The droplet radius has 
appropriately swelled, since the total number of particles is
always $N$. 

Note that (\ref{qhole}) stills respects the Gauss
constraint (\ref{AA}) (with $|v\ra = |N-1\ra$) {\it without} the
explicit introduction of any source. So, unlike the infinite
plane case, this model contains states representing quasiholes
without the need to introduce external sources. What happens
is that the hole and the boundary of the droplet together cancel
the anomaly of the commutator, the outer boundary part absorbing
an amount $N+q$ and the inner (hole) boundary producing
an amount $q$. This possibility did not exist in the infinite
plane, where the boundary at infinity was invisible,
and an explicit source was needed to nucleate the hole. We view
this as an advantage of the finite $N$ matrix model.

Quasiparticle states are a different matter. In fact, there
are {\it no} quasiparticle states with the extra particle number
localized anywhere within the droplet. Such states do not belong
to the $\nu = 1/B\theta$ Laughlin state. There are quasiparticle
states with an {\it integer} particle number $-q=m$, and the
extra $m$ electrons occupying positions {\it outside} the droplet.
The explicit form of these states is not so easy to write.
At any rate, it is interesting that the matrix model `sees' the
quantization of the particle number and the inaccessibility of
the interior of the quantum Hall state in a natural way.

Having said all that, we are now making the point that {\it
all types of states defined above are the same}. Quasihole
and quasiparticle states are nonperturbative boundary excitations
of the droplet, while perturbative boundary excitations can
be viewed as marginal particle states. 

To clarify this point,
note that the transformation (\ref{dropex}) or (\ref{newR})
defining infinitesimal boundary excitations has $2N$ real
parameters. The general state of the system, as presented in
(\ref{Xsol}) also depends on $2N$ parameters (the $x_n$ and $y_n$).
The configuration space is connected, so all states can be reached
continuously from the ground state. Therefore, all states can be
generated by exponentiating (\ref{dropex}). This is again a feature
of the finite-$N$ model: there is no sharp distinction between
`perturbative' (boundary) and `soliton' (quasiparticle) states, 
each being a particular limit of the other.

\subsection{Equivalence to the Calogero model}

We now make one of the main points of this paper.
The model examined above is, in fact, equivalent to the 
Calogero model \cite{OP,PolMM}, an integrable system of $N$ 
nonrelativistic particles on the line with hamiltonian
\be
H = \sum_{n=1}^N \left( \frac{\omega}{2B} p_n^2 + \frac{B\omega}{2}
x_n^2 \right) + \sum_{n\neq m} \frac{\nu^{-2}}{(x_n - x_m )^2}
\label{Hcalo}
\ee 
In terms of the parameters of the model, the mass of the particles
is $B/\omega$ and the coupling constant of the two-body inverse-square
potential is $\nu^{-2}$. We refer the reader to \cite{OP,PolLH,PolMM}
for details of the derivation of the connection between the matrix
model and the Calogero model. Here we simply state the relevant
results and give their connection to quantum Hall quantities.

The positions of the Calogero particles $x_n$ are the
eigenvalues of $X_1$, while the momenta $p_n$ are the diagonal
elements of $X_2$, specifically $p_n = B y_n$. The motion of the
$x_n$ generated by the hamiltonian (\ref{Hcalo}) is compatible with
the evolution of the eigenvalues of $X_1$ as it evolves in time
according to (\ref{Arot}). So the Calogero model gives a 
one-dimensional perspective of the quantum Hall state by monitoring
some effective electron coordinates along $X_1$ (the eigenvalues
of $X_1$).

The hamiltonian of the Calogero model (\ref{Hcalo}) is equal
to the potential $V = \half B\omega \Tr X_a^2$ of the matrix model.
Therefore, energy states map between the two models.
The ground state is obtained by putting the particles at their
static equilibrium positions. Because of their repulsion, they
will form a lattice of points lying at the roots of the $N$-th 
Hermite polynomial and reproducing the semi-cicle Wigner 
distribution mentioned before. 

Boundary excitations of the quantum Hall
droplet correspond to small vibrations around the equilibrium
position, that is, sound waves on the lattice. Quasiholes are
large-amplitude (nonlinear) oscillations of the particles at a
localized region of the lattice. For a quasihole of charge $q$
at the center, on the average $q$ particles near $x=0$ participate
in the oscillation. 

Finally, quasiparticles are excitations
where one of the particles is isolated outside the ground state
distribution (a `soliton') \cite{PolWS}. 
As it moves, it `hits' the distribution
on one side and causes a solitary wave of net charge 1 to 
propagate through the distribution. As the wave reaches the 
other end of the distribution another particle emerges and
gets emitted there, continuing its motion outside the distribution.
So a quasiparticle is more or less identified with a Calogero
particle, although its role, at different times, is assumed by
different Calogero particles, or even by soliton waves within the
ground state distribution.

Overall, it is amusing and useful to have the 
one-dimensional Calogero particle picture of the quantum Hall state
and translate properties back-and-forth between the two descriptions.
Further connections at the quantum level will be described in
subsequent sections.

\section{Quantization of the matrix Chern-Simons model}
\subsection{Gauss law and quantization of the filling fraction}

We now come to the question of the quantization of the above
matrix model. This is an important issue, since some of the relevant
features of the quantum Hall state will only emerge at the quantum
level. The quantization has been treated in \cite{PolMM}. 
We shall repeat here the basic arguments 
establishing their relevance to the quantum Hall system.

We shall use double brackets for quantum commutators
and double kets for quantum states, to distinguish them
from matrix commutators and $N$-vectors.

Quantum mechanically the matrix elements of $X_a$ become
operators. Since the lagrangian is first-order in time derivatives,
$(X_1)_{mn}$ and $(X_2)_{kl}$ are canonically conjugate:
\be
[[ (X_1)_{mn} , (X_2)_{kl} ]] = \frac{i}{B} \delta_{ml} \delta_{kn}
\label{Xcanon}
\ee
or, in terms of $A= X_1 + i X_2$
\be
[[ A_{mn} , A_{kl}^\dagger ]] = \frac{1}{B} \delta_{mk} \delta_{nl}
\label{Acanon}
\ee
The hamiltonian, ordered as $\half B\omega \Tr A^\dagger A$, is
\be
H = \sum_{mn} \half B \omega A_{mn}^\dagger A_{mn}
\ee
This is just $N^2$ harmonic oscillators. 
Further, the components of the vector $\Psi_n$
correspond to $N$ harmonic oscillators. Quantized as bosons, 
their canonical commutator is
\be
[[ \Psi_m , \Psi_n^\dagger ]] = \delta_{mn}
\ee

So the system is a priori just $N(N+1)$ uncoupled oscillators.
What couples the oscillators and reduces the system to effectively 
$2N$ phase space variables
(the planar coordinates of the electrons) is the Gauss law constraint
(\ref{G}). In writing it, we in principle encounter operator ordering
ambiguities. These are easily fixed, however, by noting that the
operator $G$ is the quantum generator of unitary rotations of
both $X_a$ and $\Psi$. Therefore, it must satisfy the commutation
relations of the $U(N)$ algebra. The $X$-part is an
orbital realization of $SU(N)$ on the manifold of $N \times N$
hermitian matrices. Specifically, expand $X_{1,2}$ and 
$A , A^\dagger$ in the complete basis of matrices 
$\{1, T^a\}$ where $T^a$ are the $N^2 -1$ normalized
fundamental $SU(N)$ generators:
\be
X_1 = x_0 + \sum_{a=1}^{N^2 -1} x_a T^a ~,~~~ 
\sqrt{B} A =  a_o + \sum_{a=1}^{N^2 -1} a_a T^a 
\ee
$x_a$, $a_a$ are scalar operators. Then, by 
(\ref{Xcanon},\ref{Acanon}) the
corresponding components of $B X_2$ are the conjugate operators
$-i \partial / \partial x_a$, while $a_a , a_a^\dagger$ are
harmonic oscillator operators. We can write the components of
the matrix commutator $G_X = -iB [ X_1 , X_2 ]$ in $G$ 
in the following ordering
\beqar
G_X^a &=& -i f^{abc} x_b \frac{\partial}{\partial x_a} \\
 &=& -i ( A_{mk}^\dagger A_{nk} - A_{nk}^\dagger A_{mk} ) \\
 &=& -i a_b^\dagger f^{abc} a_c 
\eeqar
where $f^{abc}$ are the structure constants of $SU(N)$.
Similarly, expressing $G_\Psi = \Psi \Psi^\dagger$ in the $SU(N)$
basis of matrices, we write its components in the ordering
\be
G_\Psi^a = \Psi_m^\dagger T_{mn}^a \Psi_n
\ee

The operators above, with the specific normal ordering, indeed 
satisfy the $SU(N)$ algebra. The expression of $G_X^a$ in terms
of $x_a$ is like an angular momentum. The expression of 
$G_\Psi^a$ in terms of the oscillators $\Psi_i$ and of
$G_X^a$ in terms of the oscillators $a_a$ is the well-known 
Jordan-Wigner realization of the $SU(N)$
algebra in the Fock space of bosonic oscillators. Specifically,
let $R_{\alpha \beta}^a$ be the matrix elements of the generators
of $SU(N)$ in any representation of dimension $d_R$, 
and $a_\alpha , a_\alpha^\dagger$
a set of $d_R$ mutually commuting oscillators. Then the operators
\be
G^a = a_\alpha^\dagger R_{\alpha \beta}^a a_\beta
\ee
satisfy the $SU(N)$ algebra. The Fock space of the oscillators
contains all the symmetric tensor products of $R$-representations
of $SU(N)$; the total number operator of the oscillators identifies
the number of $R$ components in the specific symmetric product.
The expressions for $G_\Psi^a$ and $G_X^a$ are specific cases of 
the above construction for $R^a$ the fundamental ($T^a$) or the
adjoint ($-if^a$) representation respectively.

So, the traceless part of the Gauss law (\ref{G}) becomes
\be
( G_X^a + G_\Psi^a ) |phys \ra\ra =0
\label{SUN}
\ee
where $|phys \ra\ra$ denotes the physical quantum states of the model.
The trace part, on the other hand, expresses the fact that the total
$U(1)$ charge of the model must vanish. It reads
\be
(\Psi_n^\dagger \Psi_n - NB\theta ) |phys \ra\ra = 0
\label{U1}
\ee

We are now set to derive the first nontrivial quantum mechanical
implication: the inverse-filling fraction is quantized to integer
values. To see this, first notice that the first term in (\ref{U1})
is nothing but the total number operator for the oscillators $\Psi_n$
and is obviously an integer. So we immediately conclude that $NB\theta$
must be quantized to an integer. 

However, this is not the whole story. Let us look again at the
$SU(N)$ Gauss law (\ref{SUN}). It tells us that physical states
must be in a singlet representation of $G^a$. The orbital part
$G_X^a$, however, realizes only representations arising out of
products of the adjoint, and therefore it contains only irreps
whose total number of boxes in their Young tableau is an integer 
multiple of $N$.  Alternatively, the $U(1)$ and $Z_N$ part of $U$ 
is invisible in the transformation $X_a \to U X_a U^{-1}$ and thus
the $Z_N$ charge of the operator realizing this transformation
on states must vanish. (For instance, for $N=2$,
$G^a$ is the usual orbital angular momentum in 3 dimensions which
cannot be half-integer.)

Since physical states are invariant under the sum of $G_X$ and
$G_\Psi$, the representations of $G_\Psi$ and $G_X$ must be 
conjugate to each other so that their product contain
the singlet. Therefore, the irreps of $G_\Psi$ must also have 
a number of boxes which is a multiple of $N$. 
The oscillator realization (\ref{U1}) contains all the
symmetric irreps of $SU(N)$, whose Young tableau consists of a single
row. The number of boxes equals the total number operator of the
oscillators $\Psi_n^\dagger \Psi_n$. So we conclude that $NB\theta$
must be an integer multiple of $N$ \cite{PolMM}, that is,
\be
B\theta = \frac{1}{\nu} = k ~,~~~ k={\rm integer}
\label{nuq}
\ee

The above effect has a purely group theoretic origin. The same
effect, however, can be recovered using topological considerations,
by demanding invariance of the quantum action $exp(iS)$ under gauge
$U(N)$ transformations with a nontrivial winding in the temporal
direction \cite{PolMM}. The $U(1)$ part of $U(t)$ makes a nonzero 
contribution
to the one-dimensional Chern-Simons term $\Tr A_0$, and we recover
(\ref{nuq}). This is the finite-$N$ counterpart of the level 
quantization for the noncommutative Chern-Simons term \cite{NP},
namely
\be
4\pi \lambda = {\rm integer}
\ee
By (\ref{lambda}) this is equivalent to (\ref{nuq}).

By reducing the model to the dynamics of the eigenvalues of
$X_1$ we recover a quantum Calogero model with hamiltonian
\be
H = \sum_{n=1}^N \left( \frac{\omega}{2B} p_n^2 + \frac{B\omega}{2}
x_n^2 \right) 
+ \sum_{n\neq m} \frac{k(k+1)}{(x_n - x_m )^2}
\label{Hqcalo}
\ee 
Note the shift of the coupling constant from $k^2$ to $k(k+1)$
compared to the classical case. This is a quantum reordering effect
which results in the shift of $\nu^{-1}$ from $k$ to $k+1 \equiv n$.
The above model is, in fact, perfectly well-defined even for 
fractional values of $\nu^{-1}$, while the matrix model that
generated it requires quantization. This is due to the fact that,
by embedding the particle system in the matrix model, we have
augmented its particle permutation symmetry $S_N$ to general 
$U(N)$ transformations; while the smaller symmetry $S_N$ is 
always well-defined, the larger $U(N)$ symmetry becomes anomalous 
unless $\nu^{-1}$ is quantized.

\subsection{Quantum states}

We can now examine the quantum states of this theory. The results
are, again, available and all we need to do is interpret them.

The quantum states of the model are simply states in the Fock
space of a collection of oscillators. The total energy is the
energy carried by the $N^2$ oscillators $A_{mn}$ or $a_a$.
We must also impose the constraint (\ref{SUN}) and (\ref{U1})
on the Fock states. Overall, this becomes a combinatorics
group theory problem which is in principle doable, although
quite tedious.

Fortunately, we do not need to go through it here. The quantization
of this model is known and achieves its most intuitive description
in terms of the states of the corresponding Calogero model. We
explain how.

Let us work in the $X_1$ representation, $X_2$ being its canonical
momentum. Writing $X_1 = U \Lambda_1 U^{-1}$ with $\Lambda =
{diag} \,\{x_i \}$ being its eigenvalues, we can view the state of the 
system as a wavefunction of $U$ and $x_n$. The gauge generator
$G_X^a$ appearing in the Gauss law (\ref{SUN}) is actually the
conjugate momentum to the variables $U$. Due to the Gauss law,
the angular degrees of freedom $U$ are constrained to be in a
specific angular momentum state, determined by the representation 
of $SU(N)$ carried by the $\Psi_n$.
{}From the discussion of the previous section,
we understand that this is the completely symmetric representation
with $nN = N/\nu$ boxes in the Young tableau. So the dynamics of
$U$ are completely fixed, and it suffices to consider the states
of the eigenvalues. These are described by the states of the 
quantum Calogero model. The hamiltonian of the Calogero model 
corresponds to the matrix potential $V = \half B\omega \Tr X_a^2$,
which contains all the relevant information for the system.

Calogero energy eigenstates are expressed in terms of $N$
positive, integer `quasi-occupation numbers' $n_j$ (quasinumbers,
for short), with the property
\be
n_j - n_{j-1} \geq n = \frac{1}{\nu} ~,~~~j=1,\dots N
\label{repul}
\ee
In terms of the $n_j$ the spectrum becomes identical to the
spectrum of $N$ independent harmonic oscillators
\be
E = \sum_{j=1}^N E_j = \sum_{j=1}^N \omega \left( n_j +\half\right)
\ee
The constraint (\ref{repul}) means that the $n_j$ cannot be packed 
closer than $n=\nu^{-1}$, so they have a `statistical repulsion' 
of order $n$.
For filling fraction $\nu=1$ these are ordinary fermions, while
for $\nu^{-1} = n>1$ they behave as particles with an enhanced
exclusion principle. 

The scattering phase shift between Calogero
particles is $exp(i\pi/\nu)$. So, in terms of the phase that
their wavefunction picks upon exchanging them, they look like
fermions for odd $n$ and bosons for even $n$ \cite{PolFS}. 
Since the underlying
particles (electrons) must be fermions, we should pick $n$ odd.

The energy `eigenvalues' $E_j$ are the quantum 
analogs of the eigenvalues of the matrix $\half B\omega X_a^2$.
The radial positions $R_j$ are determined by
\be
\half B\omega R_j^2 = E_j ~~~\rightarrow ~~~
R_j^2 = \frac{2 n_j +1 }{B}
\ee
So the quasinumbers $2n_j +1$ determine the radial positions 
of electrons. The ground state values are the smallest 
non-negative integers satisfying (\ref{repul})
\be
n_{j,gs} = n(j-1) ~,~~~ j=1 , \dots N
\label{gs}
\ee
They form a `Fermi sea' but with a density of states dilated
by a factor $\nu$ compared to standard fermions. This state
reproduces the circular quantum Hall droplet. Its radius maps
to the Fermi level, $R \sim \sqrt{(2n_{N,gs} +1)/B} \sim
\sqrt{2N\theta}$.

Quasiparticle and quasihole states are identified in a way 
analogous to particles and holes of a Fermi sea.
A quasiparticle state is obtained by peeling a `particle' from
the surface of the sea (quasinumber $n_{N,gs}$) and putting it to a
higher value ${n'}_N > n(N-1)$. This corresponds to an electron
in a rotationally invariant state at radial position $R' \sim
\sqrt{2({n'}_N +1)/B}$. Successive particles can be excited this
way. The particle number is obviously quantized
to an integer (the number of excited quasinumbers) and we can
only place them outside the quantum Hall droplet.

Quasiholes are somewhat subtler: they correspond to the minimal
excitations of the ground state {\it inside} the quantum Hall
droplet. This can be achieved by leaving all quasinumber $n_j$
for $j \leq k$ unchanged, and increasing all $n_j$, $j>k$ by one
\beqar
n_j &=& n(j-1) ~~~~ j \leq k \\
    &=& n(j-1)+1 ~~~ k < j \leq N 
\eeqar
This increases the gap between $n_k$ and $n_{k+1}$ to $n+1$
and creates a minimal `hole.' 

This hole has a particle number $-q = -1/n = -\nu$. 
To see it, consider removing a particle altogether from 
quasinumber $n_k$. This would create a gap of $2n$ between 
$n_{k-1}$ and $n_{k+1}$. The extra gap $n$ can be considered 
as arising out of the formation of $n$ holes (increasing $n_j$ 
for $j\geq k$ $n$ times). Thus the absence of a particle 
corresponds to $n$ holes. We therefore
obtain the important result that the quasihole charge is 
naturally quantized to units of
\be
q_h = \nu = \frac{1}{n}
\label{holeq}
\ee
in accordance with Laughlin theory.

We conclude by stressing once more that there is no fundamental
distinction between particles and holes for finite $N$. A particle
can be considered as a nonperturbative excitation of many holes
near the Fermi level, while a hole can be viewed as a coherent
state of many particles of minimal excitation.

\section{Epilogue}

We have proposed a finite matrix Chern-Simons model and presented
strong evidence that it describes a fractional quantum Hall 
droplet. We identified classical and quantum states of this 
model and related them to corresponding fractional quantum 
Hall states. The quantization of the inverse filling fraction
and, importantly, the quasihole charge quantization emerged as 
quantum mechanical consequences of this model.

The quantizations of the two parameters had a rather different
origin. We can summarize here the basic meaning of each:

Quantization of the inverse filling fraction is basically
angular momentum quantization. The matrix commutator of 
$[X_1 , X_2 ]$ is an orbital angular momentum in the compact
space of the angular parameters of the matrices, and it must
be quantized. 

Quantization of the quasihole charge is nothing but harmonic
oscillator quantization. Quasiholes are simply individual quanta
of the oscillators $A_{mn}$. The square of the radial coordinate
$R^2 = X_1^2 + X_2^2$ is basically a harmonic oscillator. 
$\sqrt{B} X_1$ and $\sqrt{B} X_2$ are canonically conjugate,
so the quanta of $R^2$ are $2/B$. Each quantum increases $R^2$
by $2/B$ and so it increases the area by $2\pi/B$. This creates
a charge deficit $q$ equal to the area times the ground state 
density $q=( 2\pi/B ) \cdot (1/2\pi\theta) = 1/\theta B = \nu$.
So the fundamental quasihole charge is $\nu$.

We further pointed out that this model is essentially equivalent
to the Calogero integrable model, providing another link between
the Calogero and Hall systems. 

The most important open question is the existence of a phase
transition of the system into a Wigner crystal at low filling
fractions. In fact, we have the complete quantum mechanics of
the model, yet we found no indication of such a phase transition.
It is plausible that the phase transition would only emerge at
the large-$N$ limit, but the (known) large-$N$ behavior of
the model does not manifest such a transition. 

A more likely 
scenario is that the dynamics of the particles will drive
such a phase transition, rather than the properties of the Laughlin
state itself. The electron interactions would presumably be an
important ingredient of the mechanism \cite{GR}. In order to probe
this question we would need to formulate a model which includes
mutual particle interactions. Such a model is presently lacking.

In \cite{Suss} is was argued that a signal for the broken
phase would be the spontaneous breaking of the $X$-space
area-preserving diffeomorphism invariance. The vacuum of the
infinite plane Chern-Simons theory is invariant under all
these transformations, since they amount to canonical 
transformations of the Heisenberg algebra 
$[ X_1 , X_2 ] = i \theta$ which are gauge transformations.
The matrix model proposed here breaks this invariance, as
we pointed out in section (3.3). This is due, however, to
the finite extent of the model, and specifically to the
presence of the boundary of the droplet, and it is not
an indication of a phase transition.

Although the above model gives an accurate and compelling
description of the fractional quantum Hall states and many of
their properties, it still leaves unanswered some questions
that could in principle be addressed in the quantum mechanical
many-body language. In particular, we do not have a good description
for quantities like the electron density and correlation functions.
Clearly the dictionary between the two systems needs to be 
expanded and completed. 
Functional descriptions of the quantum phase space (viewed as a
noncommutative space) such as
Wigner functions (see, e.g., \cite{CFZU}) may prove useful tools.
The above questions are left for future research.

\vskip 0.2in
{\it Acknowledgement}: I would like to thank V.P.~Nair 
for interesting discussions.


\vskip 0.2in



\begin{thebibliography}{99}


\bibitem{DJT}
S.~Deser, R.~Jackiw and S.~Templeton,
Phys.\ Rev.\ Lett.\ {\bf 48}, 975 (1982)
and Annals Phys.\ {\bf 140}, 372 (1982).

\bibitem{CF}
A.~H.~Chamseddine and J.~Frohlich,
J.\ Math.\ Phys.\ {\bf 35}, 5195 (1994)
[hep-th/9406013].

\bibitem{Kr}
T.~Krajewski, math-ph/9810015.

\bibitem{PolCS}
A.~P.~Polychronakos,
JHEP{\bf 0011}, 008 (2000)
[hep-th/0010264].

\bibitem{Chu}
C.~Chu, Nucl.\ Phys.\ B {\bf 580}, 352 (2000)
[hep-th/0003007].

\bibitem{CW}
G.~Chen and Y.~Wu,
Nucl.\ Phys.\ B {\bf 593}, 562 (2001)
[hep-th/0006114].

\bibitem{MS}
S.~Mukhi and N.~V.~Suryanarayana,
JHEP{\bf 0011}, 006 (2000)
[hep-th/0009101].

\bibitem{GS}
N.~Grandi and G.~A.~Silva,
hep-th/0010113.

\bibitem{LMS}
G.~S.~Lozano, E.~F.~Moreno and F.~A.~Schaposnik,
JHEP{\bf 0102}, 036 (2001)
[hep-th/0012266].

\bibitem{KP}
A.~Khare and M.~B.~Paranjape, hep-th/0102016.

\bibitem{BKSY}
D.~Bak, S.~K.~Kim, K.~Soh and J.~H.~Yee,
hep-th/0102137.

\bibitem{DuJT}
G.~Dunne, R.~Jackiw and C.~Trugenberger,
Phys.\ Rev.\ D {\bf 41}, 661 (1990);
G.~Dunne and R.~Jackiw,
Nucl.\ Phys.\ Proc.\ Suppl.\ {\bf 33C}, 114 (1993)
[hep-th/9204057].

\bibitem{Suss}
L.~Susskind, hep-th/0101029.

\bibitem{Laug}
R.~B.~Laughlin, `The Quantum Hall Effect,' R.~E.~Prange and
S.~M.~Girvin (Eds), p. 233.

\bibitem{BST}
J.~H.~Brodie, L.~Susskind and N.~Toumbas,
JHEP{\bf 0102}, 003 (2001) [hep-th/0010105].

\bibitem{BN}
I.~Bena and A.~Nudelman,
JHEP{\bf 0012}, 017 (2000) [hep-th/0011155].

\bibitem{GR}
S.~S.~Gubser and M.~Rangamani, hep-th/0012155.

\bibitem{CCMM}
L.~Capiello, G.~Cristofano, G.~Maiella and V.~ Marotta,
hep-th/0101033.

\bibitem{NP}
V.~P.~Nair and A.~P.~Polychronakos, hep-th/0102181.

\bibitem{BLP}
D.~Bak, K.~Lee and J.-H.~Park, hep-th/0102188
\bibitem{Cal}

F.~Calogero,
J.\ Math.\ Phys.\ {\bf 12}, 419 (1971).

\bibitem{Sut}
B.~Sutherland,
Phys.\ Rev.\ A {\bf 4}, 2019 (1971)
and Phys.\ Rev.\ A {\bf 5}, 1372 (1972).

\bibitem{Mos}
J.~Moser, Adv.\ Math.\ {\bf 16} (1975) 1.

\bibitem{OP}
M.~A.~Olshanetsky and A.~M.~Perelomov,
Phys.\ Rept.\ {\bf 71} (1981) 313
and {\bf 94} (1983) 6.

\bibitem{PolLH}
For a review closest in spirit to the present discussion,
see A.~P.~Polychronakos,
``Generalized statistics in one dimension,'' published in
`Topological aspects of low-dimensional systems,' Les Houches
Session LXIX (1998), Springer Ed. [hep-th/9902157].

\bibitem{PolFS}
A.~P.~Polychronakos,
Nucl.\ Phys.\ B {\bf 324}, 597 (1989).

\bibitem{LM}
J.~M.~Leinaas and J.~Myrheim, 
Phys.\ Rev.\ {\bf B37} (1988) 9286.

\bibitem{PolAN}
A.~P.~Polychronakos,
Phys.\ Lett.\ B {\bf 264}, 362 (1991).

\bibitem{BHKV}
L.~Brink, T.~H.~Hansson, S.~Konstein and M.~A.~Vasiliev,
Nucl.\ Phys.\ B {\bf 401}, 591 (1993)
[hep-th/9302023].

\bibitem{IR}
S.~Iso and S.~J.~Rey,
Phys.\ Lett.\ B {\bf 352}, 111 (1995)
[hep-th/9406192].

\bibitem{Ouv}
S.~Ouvry, cond-mat/9907239.

\bibitem{GMS}
R.~Gopakumar, S.~Minwalla and A.~Strominger,
JHEP{\bf 0005}, 020 (2000) [hep-th/0003160].

\bibitem{PolFT}
A.~P.~Polychronakos,
Phys.\ Lett.\ B {\bf 495}, 407 (2000)
[hep-th/0007043].

\bibitem{GN}
D.~J.~Gross and N.~A.~Nekrasov,
JHEP{\bf 0010}, 021 (2000) [hep-th/0007204]
and hep-th/0010090.

\bibitem{HKL}
J.~A.~Harvey, P.~Kraus and F.~Larsen,
JHEP{\bf 0012}, 024 (2000) [hep-th/0010060].

\bibitem{PolMM}
A.~P.~Polychronakos,
Phys.\ Lett.\ B {\bf 266}, 29 (1991).

\bibitem{Wen}
X.~G.~Wen,
Phys.\ Rev.\ B {\bf 41}, 12838 (1990).

\bibitem{IKS}
S.~Iso, D.~Karabali and B.~Sakita,
Nucl.\ Phys.\ B {\bf 388}, 700 (1992)
[hep-th/9202012]
and Phys.\ Lett.\ B {\bf 296}, 143 (1992)
[hep-th/9209003].

\bibitem{CTZ}
A.~Cappelli, C.~A.~Trugenberger and G.~R.~Zemba,
Nucl.\ Phys.\ B {\bf 396}, 465 (1993)
[hep-th/9206027].

\bibitem{PolWS}
A.~P.~Polychronakos,
Phys.\ Rev.\ Lett.\ {\bf 74}, 5153 (1995)
[hep-th/9411054].

\bibitem{CFZU}
T.~Curtright, D.~Fairlie and C.~Zachos,
Phys.\ Rev.\ D {\bf 58}, 025002 (1998)
[hep-th/9711183];
T.~Curtright, T.~Uematsu and C.~Zachos, hep-th/0011137.


\end{thebibliography}
\end{document}